# Theoretical realization of hybrid Weyl state and associated high catalytic performance for hydrogen evolution in NiSi

Wei Liu,[1,2] Xiaoming Zhang[1,2*], Weizhen Meng,[1,2] Ying Liu,[2] Xuefang Dai,[2] and Guodong Liu[1,2*]

[1]State Key Laboratory of Reliability and Intelligence of Electrical Equipment, Hebei University of Technology, Tianjin 300130, China

[2]School of Materials Science and Engineering, Hebei University of Technology, Tianjin 300130, China

Correspondence: zhangxiaoming87@hebut.edu.cn; gdliu1978@126.com

**SUMMARY:**

For electrochemical hydrogen evolution reaction (HER), developing high-performance catalysts without containing precious metals has been a major research focus in the current. Herein, we show the feasibility of HER catalytic enhancement in Ni-based materials based on topological engineering from hybrid Weyl states. Via a high-throughput computational screening from ~140 000 materials, we identify a chiral compound NiSi is a hybrid Weyl semimetal (WSM) with showing bulk type-I and type-II Weyl nodes and long surface Fermi arcs near the Fermi level. Sufficient evidences verify that topological charge carriers participate in the HER process, and make the certain surface of NiSi highly active with the Gibbs free energy nearly zero (0.07 eV), which is even lower than Pt and locates on the top of the volcano plots. This work opens up a new routine to develop no-precious-metal-containing HER catalysts via topological engineering, rather than traditional defect engineering, doping engineering, or strain engineering.

**INTRODUCTION**

It has seen rapid global energy consumption because of explosion of urbanization and industrialization in today's world. During the production of energy, the overdependence on fossil fuels has induce overmuch carbon emission and the associated environmental degradation.[1] Facing on these issues, much effort has been made on developing green energy carriers.[2,3] Among them, hydrogen has been hoped for the most as the alternative carrier to fossil fuels for future electric powering.[4-6] Till now, various methods have been proposed for hydrogen production, of which water electrolysis is viewed as an effective approach.[7-10] Fundamental mechanism on electrochemical hydrogen evolution reaction (HER) has already been well demonstrated, and has also played an important platform to understand other heterogeneous reactions.[6,11,12] Especially, the well-known volcano plots have been set up, which can soundly describe the surface catalytic activity in transition-metal-based catalysts for HER process. On volcano plots, precious-metal Pt nearly locates on the mountaintop with hosting the most profitable Gibbs free energy ($\Delta G_{H*}$).[13-15] However, the utilization of Pt is largely limited by its significantly high cost for large-scale hydrogen production. For this reason, exploring high-efficiency catalysts based on cheaper transition-metals such as Ni are highly desirable. Nevertheless, Ni fundamentally has a larger ︱$\Delta G_{H*}$︱ than Pt (0.22 eV versus 0.09 eV).[14] Various approaches such as chemical doping, phase engineering and defect engineering have been tried to speed up the surface catalytic activity in Ni.[16-19] Much progress is obtained under these attempts, but Ni-based catalysts still have much disparity comparing with Pt.

Since the establishment of topological framework, the field of condensed matter physics has undergone a revolutionary innovation. Starting from the earliest reported topological insulators,[20,21] topological semimetals with Weyl/Driac nodes, various nodal lines and nodal surface have gradually become the focus of current research.[22-29] The application of topological materials has been rapidly expanded due to their unique physical properties and electronic characteristic.[30,31] Especially, it has been demonstrated recently that metallic surface states in topological materials could favor heterogeneous catalysis.[32,33] Such metallic surface states for catalysis speed-up includes

Dirac ones in topological insulator, Fermi arc ones in Weyl/Dirac semimetals and drumhead ones in nodal line semimetals, as evidenced in tens of topological materials by both theoretical and experimental investigations.[33-41] Such phenomenon inspired us with the following question: can the Ni-based materials be combined with the novel physical features related to the topologically nontrivial surface states to enhance the activity in HER process?

To pursue this question, herein, we carry out topological engineering in Ni-based materials toward high-efficiency HER catalysts. We first build up the designing scheme, which highlights hybrid Weyl semimetals (WSMs) with different types of Weyl nodes and long surface Fermi arcs are the ideal material platform for the target. Then we perform high-throughput material screening to explore concrete examples for Ni-based hybrid WSMs. Further, the identified material NiSi is theoretically demonstrated to show hybrid Weyl nodes, long surface Fermi arcs and a nearly zero $\Delta G_{H^*}$, suggesting excellent prospect as high-efficiency HER catalysts. The nontrivial band topology and the associated long Fermi arcs on the surface are evidenced to be responsible for the high HER activity in NiSi. This work highly promises topological engineering be a feasible approach to realize high performance catalysts without involving precious metals.

**DESINGING SCHEME**

Among various topological HER catalysts proposed previously,[33-41] WSMs have been one of the most remarkable categories, where the surface Fermi arcs originating from bulk Wey nodes are believed to enhance catalytic activity by providing robust active center, high carrier density and mobility around the Fermi energy.[35] The situation seems wonderful! Unfortunately, it seems that the catalytic enhancement in existing WSMs only works in a finite level, which largely because that the length of Fermi arcs in traditional WSMs is usually very short. Thus, this designing scheme aims at developing WSMs with long surface Fermi arcs.

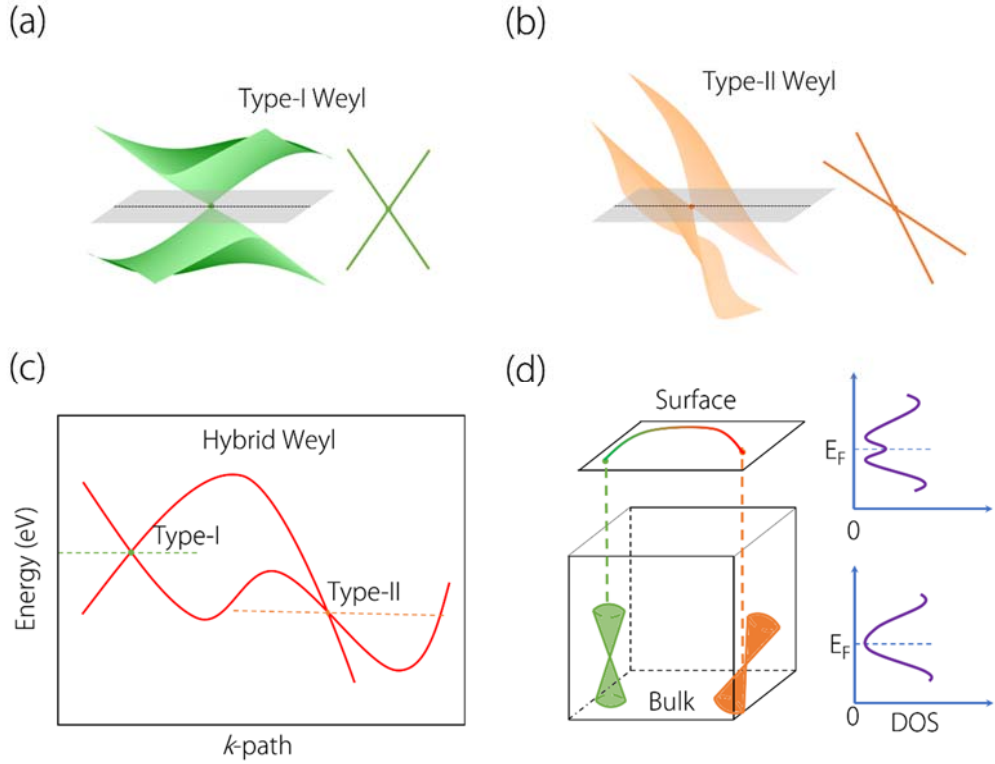

Figure 1 3D and 2D band dispersion model for (a) type-I and (b) type-II Weyl point. (c) Illustration of hybrid nodal points composed by one type-I and one type-II Weyl point. (d) Schematic of hybrid Weyl points in the bulk (left lower panel) and associated long Fermi arc on the surface (left upper panel). The DOSs for the bulk Weyl points and surface Fermi arc are provided in right panels.

Then, how to realize long Fermi arcs in WSMs? Considering the tilting degree of the Weyl cone in the momentum space, Weyl nodes can be classified as two typical categories (i.e. type-I and type-II).[42,43] As shown in Fig. 1(a) and (b), type-I and type-II Weyl nodes show conventional and totally tilted band crossing, respectively. For WSMs contains only one type (either type-I or type-II) of Weyl nodes, long surface Fermi arcs are less likely to appear because the pairs of Weyl nodes with opposite charities are axisymmetrical distributed (assuming that the system preserves the time-reversal symmetry and only contains few pairs of Weyl nodes). However, the situation will change if type-I and type-II of Weyl nodes coexist in a single WSM, as shown by the case in Fig. 1(c). In this case, type-I and type-II of Weyl nodes happen in different $k$-paths (also usually locating at different energy levels), thus the Fermi arcs which connect these Weyl nodes in principle need to span a large area on the surface. In fact, such novel WSM with hosting both type-I and type-II Weyl nodes has already proposed in model in 2016, defined as hybrid WSM.[44] Figure 1(d) shows the catalytic mechanism for a hybrid WSM. The figure has indicated the presences of type-I and type-II Weyl nodes in the bulk, which generate long Fermi arc on the surface and the associated high surface electron density of states (DOSs).

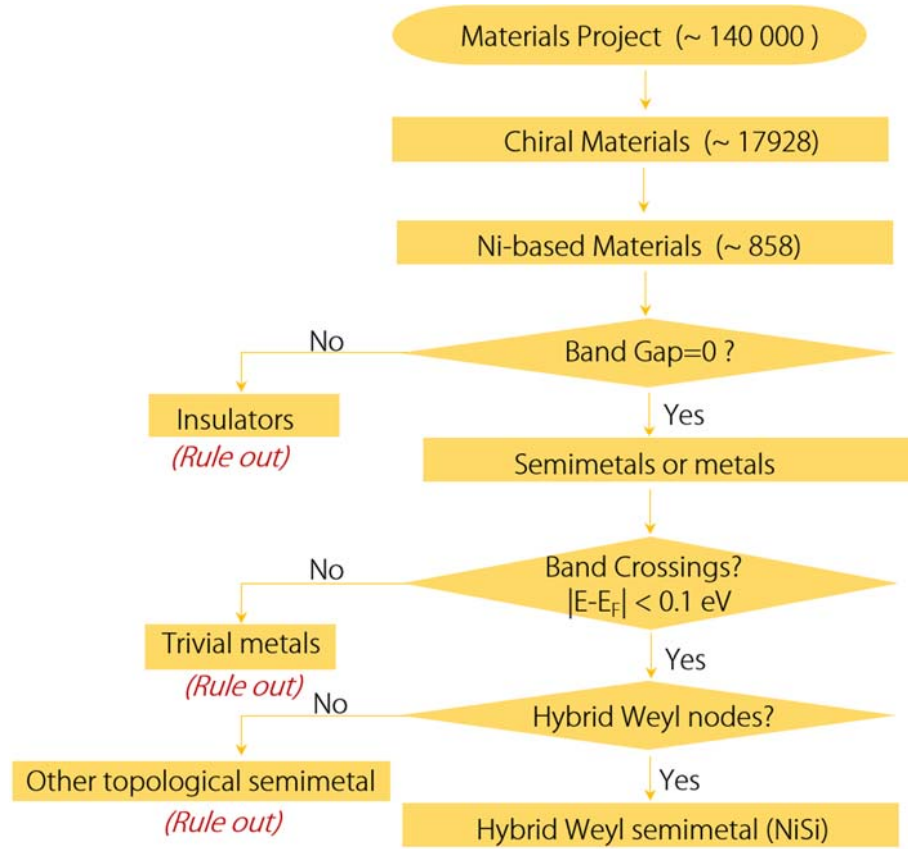

Figure 2 Flowchart of high-throughput computational screening for Ni-based hybrid Weyl semimetal from the Materials project database.

To testify above assumption, the most crucial determination lies in identifying hybrid WSM candidate in Ni-based compounds. This is a challenging task, noticing that excellent hybrid WSMs are quite rare, which were only identified in few materials including $T_d$-MoTe$_2$, OsC$_2$, YCoC$_2$ and HfCuP.[45-49] On considering this fact, we make a high-throughput material screening based on Material Project[50] (with ~140 000 entries). We show the computational screening process in Fig. 2. For the first step, we screen out the materials under chiral space groups from Materials Project data, after which the number of candidates has been sharped down to 17 928. Then, we exclude chiral compounds without containing Ni element, which results in 858 compounds. Further, we exclude insulating chiral Ni-based compounds by assessing band gap, and metallic ones are retained for further computation. For the next step, we carry out band structure calculations for detecting potential band crossings, which are essential for WSMs. Only the band crossings within $|E - E_F| < 0.1$ eV are taken into account, because the electrons in this region are the most likely to contribute the conducting natural of the material. Furthermore, we identify the nature of the band crossings, and the final candidate materials need to contain both type-I and type-II Weyl nodes. During material screening, we fortunately find compound NiSi can satisfy all the rigorous screening standards. Thus, in the following, we systematically investigate the topological feature and catalytic performance for HER.

**METHODS**

The work is performed under numerical calculations based on the density function theory (DFT),[51] as realized via the Vienna ab initio simulation package (VASP).[52] As an inexpensive generalized gradient approximation (GGA), the exchange-related Potential- Burke-Ernzerhof (PBE)[53] function is adopted in current work. The cutoff energy is set to be 500eV. For the calculations for electronic band structure of NiSi, the bulk Brillouin zone (BZ) is sampled with the *k*-mesh of $9 \times 9 \times 9$ with the Γ point as the center. During calculation, the energy and force convergence standards are set to be $10^{-9}$ eV and 0.01 eV/Å$^{-1}$, respectively. The topologically surface status is calculated based on treatments as implemented in the WANNIERTOOLS package.[54]

For the surface model for H adatom adsorption, we have tested slab thickness (*n*) by examining the surface energy dispersion with gradually increasing the number of *n*. The simulation shows the energy would show no dispersion with n ⩾ 5 during further add the slab thickness. Considering this, the surface model for NiSi is constructed by using a 10 unit-cell-thick slab (n= 10) with a 2 ×2 bottom supercell immobilized. In addition, during the simulation the DFT-D2 extension of Grimme is used to account for the long-range van der Waals interactions.[55] To explore the HER activity in NiSi compound, we calculate the value of $\Delta G_{H*}$ following the formula:[56]

$$\Delta G_{H*} = \Delta E_H + \Delta E_{ZPE} - T \Delta S_H \quad (1)$$

where $\Delta E_H$ is the adsorption energy of single H, $\Delta E_{ZPE}$ ($\Delta S_H$) denotes the difference in zero-point energy (entropy) between the absorbed H and gaseous H.

**RESULTS AND DISCUSSIONS**

**Crystal structure and band topology of NiSi**

For Ni-Si compound with the 1:1 stoichiometry, various phases under orthorhombic, tetragonal, and also cubic ones were reported to be stable in literatures.[57-65] The NiSi compound focused in this work is a cubic phase taking the ε-FeSi (P213, B20) structure. The structure belongs the chiral space group *P213* (No.198). This NiSi phase has been successfully synthesized by different methods previously.[57,63-65] The atomic lattice structure of NiSi is shown in Fig. 3(a). One NiSi unit cell contains 4 Ni and 4 Si atoms. The atoms are arranged in the distorted face-centered cubic structure, where the Ni atoms are forced out from their original positions by the Si atoms. The optimized lattice constant for NiSi is $a = b = c = 4.496$ Å, being comparable with experimental one (4.437 Å [57]).

The band structure and the electronic density of states (DOSs) of NiSi are displayed in Fig. 3(c). Two Weyl nodes (remarked as $W_1$ and $W_2$) can be observed near the Femi level, where $W_1$ situates at 0.079 eV above the Fermi level locating on the *k*-path R-Γ while $W_1$ locates at 0.005 eV on the Γ-X path. The provided DOSs indicates the conducting electronic states are almost contributed by the 3*d* orbitals of the Ni atom. The Weyl nodes manifest different band slops where $W_1$ is a type-I Weyl node while $W_2$ is type-II. These Weyl nodes are formed by the intersections between the same two bands, which can be clearly observed by the 3D plotting of the bands shown in Fig. 3(d). Thus, NiSi compound can be well characterized as a hybrid WSM. We take the (001) surface projection [see Fig. 3(b)] to investigate the surface states of NiSi compound. We show the slice of surface states at the Fermi energy in Fig. 3(e). Considering the $C_4$ symmetry in the system, the slice range is only selected as 1/4 of the whole surface BZ. We indeed observe long Fermi arcs corresponding to Weyl nodes $W_1$ and $W_2$, where the extension of Fermi arcs spans most region of the slice, as shown in Fig. 3(c).

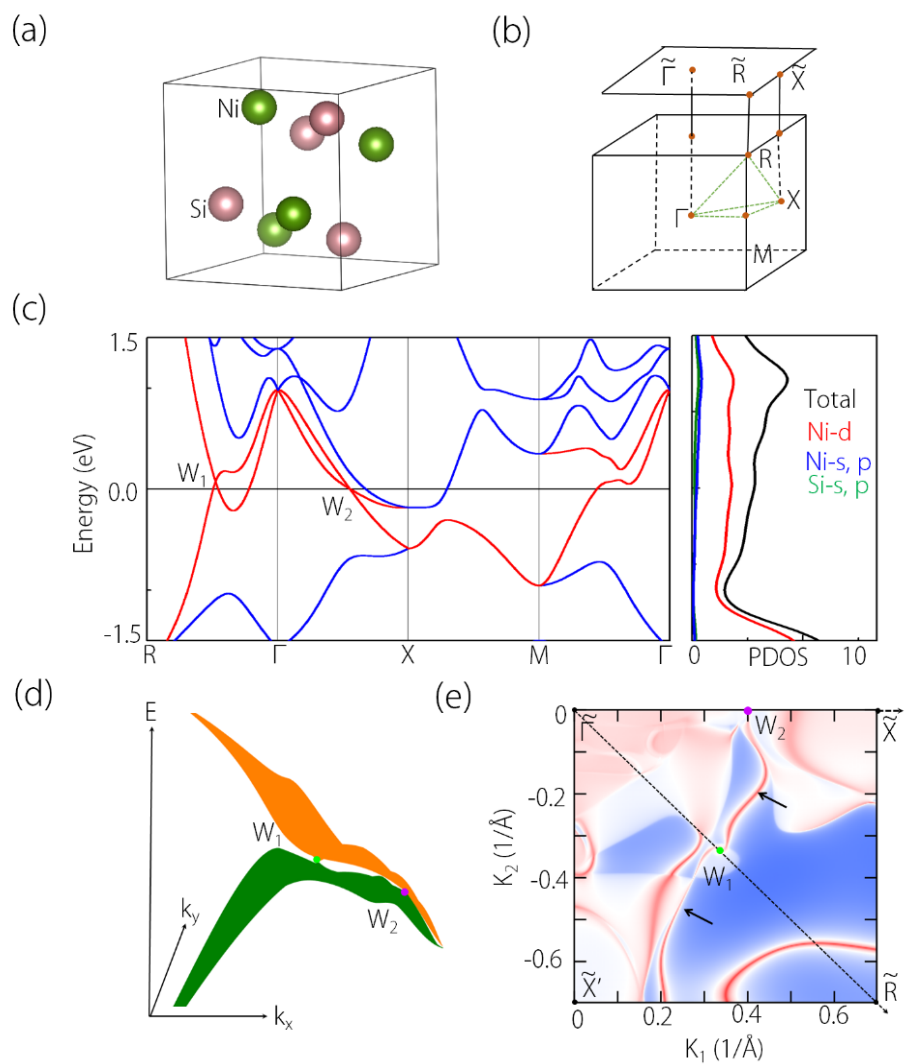

Figure 3 (a) Atomic lattice structure of NiSi compound. (b) The bulk and the (001) surface Brillouin zone with high-symmetry *k*-points provided. (c) Electronic band structure projected density of states (PDOS) of NiSi compound. The two band crossings in the band structure are denoted as $W_1$ and $W_2$. (d) 3D plotting of band dispersions for the hybrid Weyl nodes. (e) The constant energy slices corresponding to (010) surface at the Fermi level. The Fermi arcs are pointed by the arrows.

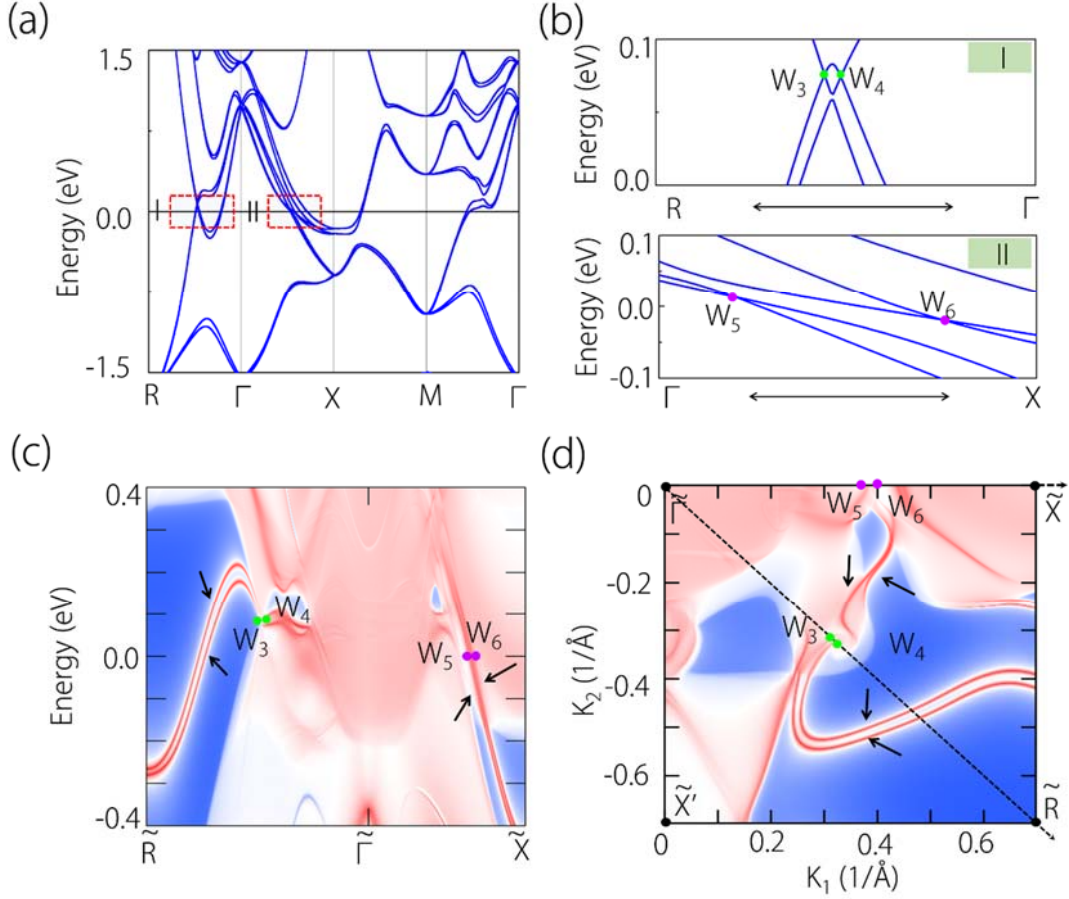

Figure 4 (a) Electronic band structure of NiSi compound with SOC included. The band crossings near the Fermi level are highlighted by two regions (I and II in the figure). (b) Enlarged band structure for region I and region II. The four Weyl points are denoted as $W_3$-$W_6$. (c) Projected spectrum on the (001) surface of NiSi. The positions of the projected hybrid Weyl points and Fermi arcs are shown by the dot and arrow, respectively. (d) The constant energy slices corresponding to (010) surface at the Fermi level.

The spin-orbit coupling (SOC) cannot be simply ignored in NiSi. Thus, in the following we discuss the electronic band with taking into account SOC. As shown in Fig. 4(a), we find the bands experience slight split under SOC. Still, we pay special attention to the bands in the R-Γ and Γ-X paths (remarked as region I and II). The enlarged band structures for them are shown in Fig. 4(b). We find the Weyl nodes in NiSi is multiplied under SOC, where two type-I Weyl nodes ($W_3$ and $W_4$) appear in the R-Γ path and two type-II ones ($W_5$ and $W_6$) in Γ-X path. These results indicate NiSi is still a hybrid WSM under SOC. The (001) surface band structure is shown in Fig. 4(c), where the Fermi arcs from the Weyl nodes can be clearly identified. Furthermore, we display the slice of surface states at the Fermi energy in Fig. 4 (d). We find these long Fermi arcs almost transverse the entire slice space.

**HER performance of NiSi**

With the hybrid Weyl natural and the existence of long surface Fermi arcs in minds, we continue to investigate the HER catalytic performance in NiSi compound. The possible HER catalytic mechanism in NiSi is schematically shown in Fig. 5(a). The hybrid Weyl nodes bring long Fermi arcs on the (001) surface of NiSi, which makes the surface with high activity and spires the HER process by reducing the $\Delta G_{H*}$ and increasing the surface conductivity. We simulate the HER process by adsorbing single H atom on the NiSi slab with the Ni-terminated (001) surface. Our calculations show the most favorable adsorption site for H is on the top of Ni atom. Under this state, we provide the map of the charge density difference in Fig. 5(b). We find sizable charge depletion on the (001) surface of NiSi (see the left panel), which has transferred to H adatom (i.e. charge accumulation on H, see the right panel). In addition, in Fig. 5(c) and (d) we compare the surface band structure of NiSi before and after H adsorption. It can be clearly seen that the four Weyl node points move down after adding H atoms on the NiSi surface, indicating that the charge transfer takes place on its surface. We find the Fermi arcs moves toward lower energy after H adsorption, further verifying the charge exchange between NiSi surface and H atom during the HER process.

The HER activity in NiSi can be directly reflected by $\Delta G_{H*}$. Just as expected, the calculated $\Delta G_{H*}$ is close to zero (0.077 eV). In Fig. 5 (e), we compare the HER activity between NiSi and Weyl HER catalysts proposed before. For comparison, the case of pure Ni is also provided. We find ∣$\Delta G_{H*}$∣ of NiSi (0.07 eV) is significantly lower than WSMs NbAs (0.96 eV), TaAs (0.74 eV), TaP (0.38 eV), and NbP (0.31 eV),[35] and Ni (0.22 eV) as well.[14] The relatively low $\Delta G_{H*}$ benefits from the long surface Fermi arcs in NiSi. Furthermore, we show the volcano curves in Fig. 5(f). We can observe that NiSi almost locates at the top of the volcano curves. Remarkably, by only view on the $\Delta G_{H^*}$, the HER activity in NiSi is even better than the precious Pt (0.07 eV *v.s.* 0.09 eV[14]). These results suggest the hybrid WSM NiSi is promising to be a superior HER catalyst without containing precious-metal elements.

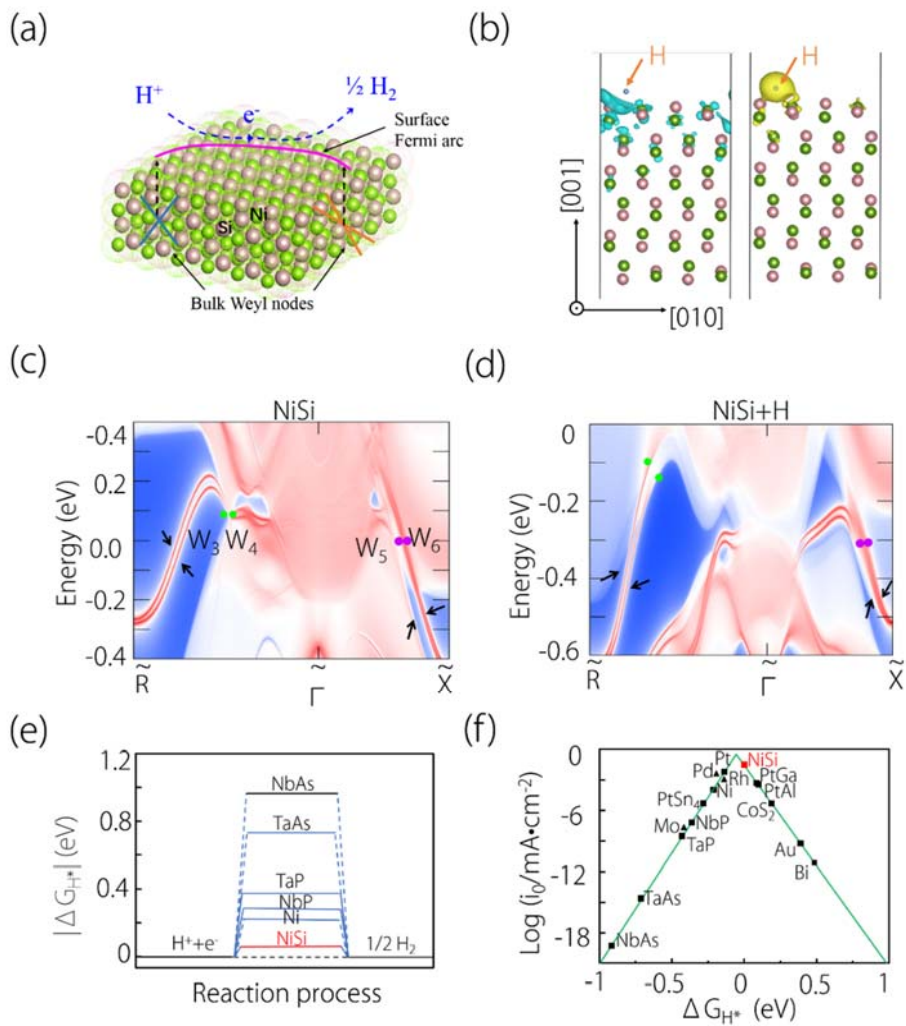

Figure 5 (a) Schematic illustration of the HER process on the (001) surface with the enhanced activity by the hybrid Weyl nodes induced long Fermi arcs. (b)The electron depletion (the left pannel) and accumulation (the right panel) during H adsorption on NiSi (001) surface. (c) and (d) show the (001) surface states of NiSi before and after the hydrogen adsorption, respectively. (e) The free energy diagram for hydrogen evolution for NiSi and typical Weyl catalysts. The case of Ni is also provided for comparison. The data for NbAs, TaAs, NbP and TaP are taken from ref. 81. The data for PtSn$_4$ is taken from ref. 36. The data for Ni is taken from ref. 15. (f) Volcano plot for HER of NiSi in comparison with typical catalysts. Except NiSi, the data for other catalysts are taken from refs. 15, 34-42.

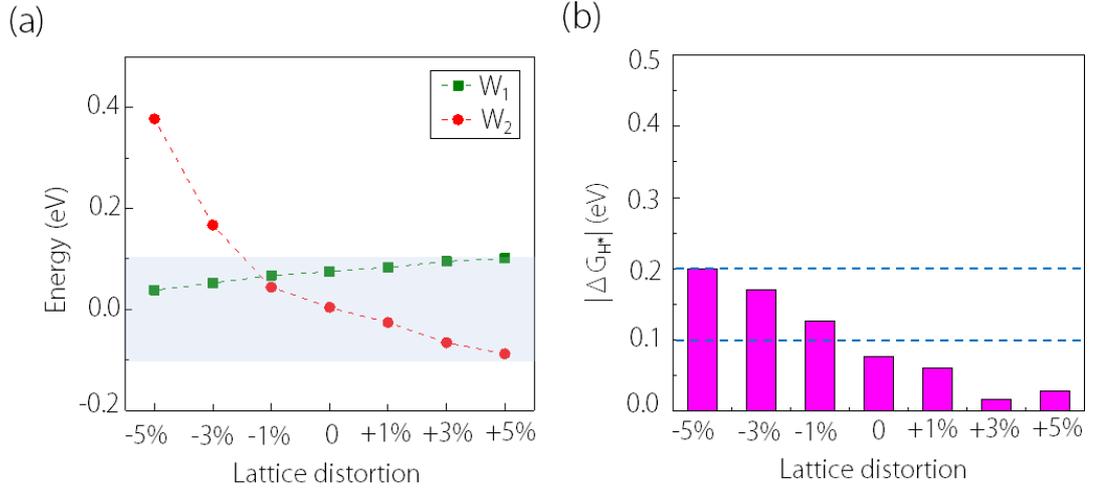

Figure 6 (a) The positions of the hybrid Weyl nodes ($W_1$ and $W_2$) under hydrostatic lattice distortions from -5% to +5%, where negative and positive values represent lattice compression and lattice expansion, respectively. (b) The calculated $|\Delta G_{H*}|$ on the (001) surface of NiSi under hydrostatic lattice distortions from -5% to +5%.

To fully ascertain the relationship between the hybrid Weyl nodes and the HER activity, we artificially tune the positions of the hybrid Weyl nodes by different approaches. As the first, we perform hydrostatic distortions to the NiSi lattice. In Fig. 6(a), we show the positions of the hybrid Weyl nodes ($W_1$ and $W_2$) under hydrostatic lattice distortions from -5% to +5%, where negative and positive values represent lattice compression and lattice expansion, respectively. One can find that the position of $W_1$ does not change much during lattice distortions, which almost retains in the energy range of ± 0.1 eV. For comparison, the position of $W_2$ experiences greater changes, especially under lattice compression. Notably, as shown by the shadowed region in Fig. 6(a), both $W_1$ and $W_2$ locate quite near the Fermi level (within $|E - E_F| < 0.1$ eV) at most lattice distortions. This indicates hybrid Weyl nodes and the surface Fermi arcs in NiSi are weakly sensitive to the lattice distortions. Correspondingly, the high (001) surface HER activity is expected to retain as well. This expectation has been verified by our calculations. As shown in Fig. 6(b), the calculated $|\Delta G_{H*}|$ on the (001) surface for all the cases are below 0.2 eV, suggesting high HER activity in NiSi. These results suggest the HER activity is highly relative to the hybrid Weyl nodes in NiSi.

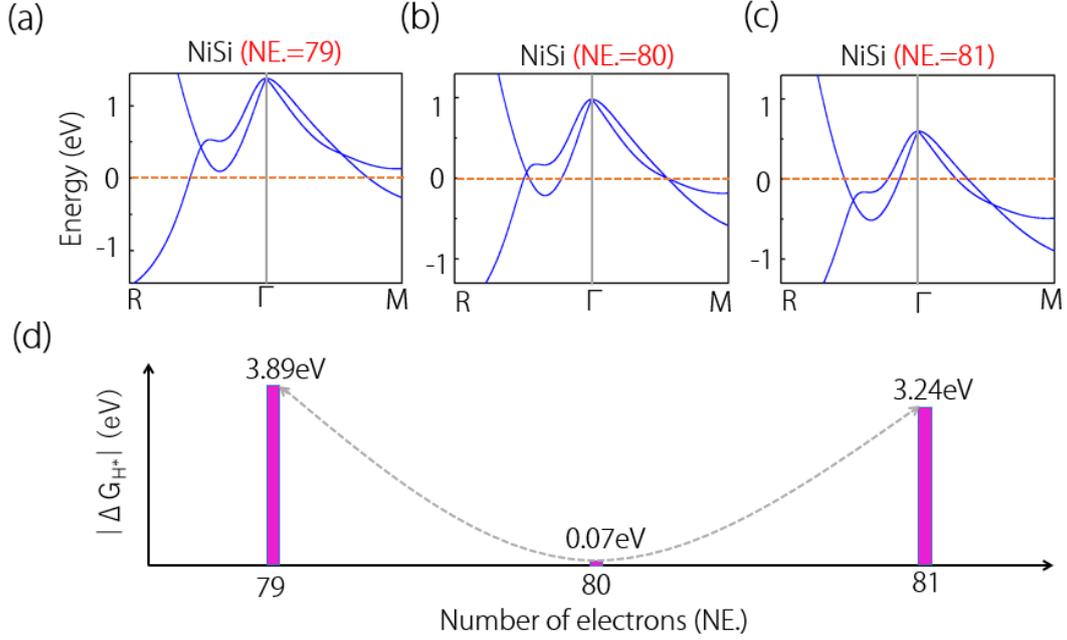

Figure 7 Electronic band structure of NiSi with losing one electron. The native number of electrons (NE.) in a NiSi cell is 80. (b) and (c) are the band structures for native NiSi (NE. = 80) and that with gaining one electron (NE. = 81), respectively. (d) Comparison of ︱$\Delta G_{H*}$︱ in NiSi among NE.= 79, 80, and 81.

Since the hybrid Weyl nodes are weakly sensitive to lattice distortions, here we further tune the hybrid Weyl nodes by shifting the number of electrons (NE.) in NiSi system. As shown by the band structure in Fig. 7(a), both the type-I and type-II Weyl nodes will move to high energy levels (at about 0.3 eV to 0.4 eV) when NiSi system loses one electron. On the contrary, the Weyl nodes will be pulled into low energy levels (at about -0.3 eV) when NiSi system gains one electron, as shown in Fig. 7(c). For comparison, the band structure for native NiSi is provided in Fig. 7(b), where the Weyl nodes locate quite near the Fermi level ($W_1$ at 0.079 eV and $W_2$ at 0.005 eV). Comparing with native NiSi (NE.= 80), the hybrid Weyl nodes and corresponding surface Fermi arcs for the cases with losing one electron (NE.= 79) and gaining one electron (NE.= 81) are expected to be less contributing to the HER process, because the Weyl nodes in these cases are far away from the Fermi level. In Fig. 7(d), we compare the ︱$\Delta G_{H*}$︱ on the (001) surface of NiSi with NE.= 79, 80, and 81. The results show that, with tuning the Weyl nodes away from the Fermi level, ︱$\Delta G_{H*}$︱ for NiSi system indeed greatly increase (3.89 eV for Nelec.= 79, and 3.24 eV for Nelec.= 81), suggesting the weakness of HER activity during the period. These results have fully revealed the strong relation between HER activity and hybrid Weyl nodes in NiSi.

**Conclusion**

In conclusion, we have demonstrated hybrid WSM is a promising platform to enhance catalytic activity for HER, where the long Fermi arc states related to hybrid Weyl nodes make certain surface highly active. We perform high-throughput material screening from ~140 000 materials within the Material Project database, and fortunately identify a Ni-based compound namely NiSi as a new hybrid WSM with high HER activity. NiSi follows a chiral space group, and manifests both type-I and type-II Weyl nodes near the Fermi level. We further verify that on its surface, where long Fermi arcs from hybrid Weyl nodes present, $\Delta G_{H^*}$ for HER process is almost zero (0.07 eV), being favorable for high-efficiency catalyst. Remarkably, the absolute value of $\Delta G_{H^*}$ in NiSi is even smaller than Pt (0.07 eV *v.s.* 0.09 eV), and NiSi almost locates on the top of the volcano curves. This work provides new insight on developing high-efficiency catalyst without precious metals by using hybrid WSMs.

**ACKNOWLEDGEMENTS**

This work is supported by National Natural Science Foundation of China (Grants No. 11904074). The work is funded by Science and Technology Project of Hebei Education Department, the Nature Science Foundation of Hebei Province, S&T Program of Hebei (A2019202107), the Overseas Scientists Sponsorship Program by Hebei Province (C20200319 and C20210330). The work is also supported the State Key Laboratory of Reliability and Intelligence of Electrical Equipment (No. EERI_OY2020001), Hebei University of Technology. One of the authors (X.M. Zhang) acknowledges the financial support from Young Elite Scientists Sponsorship Program by Tianjin.

**DECLARATION OF INTERESTS**

The authors declare no competing financial interests.